\documentclass[%
 reprint,
superscriptaddress,
 amsmath,amssymb,
 aps,
prb,
]{revtex4-2}

\usepackage{graphicx}
\usepackage{dcolumn}
\usepackage{bm}
\usepackage{siunitx}
\usepackage{amsmath}
\usepackage{mathtools}

\begin{document}

\preprint{APS/123-QED}

\title{Impact of surface anisotropy on the spin-wave dynamics in thin ferromagnetic film}

 
\author{Krzysztof~Szulc}
\email{krzysztof.szulc@amu.edu.pl}
\affiliation{%
Institute of Spintronics and Quantum Information, Faculty of Physics, Adam Mickiewicz University, Pozna\'{n}, Uniwersytetu Pozna\'{n}skiego 2, 61-614 Pozna\'{n}, Poland 
}%
\author{Julia Kharlan}%
\affiliation{%
Institute of Spintronics and Quantum Information, Faculty of Physics, Adam Mickiewicz University, Pozna\'{n}, Uniwersytetu Pozna\'{n}skiego 2, 61-614 Pozna\'{n}, Poland 
}%
\affiliation{%
Institute of Magnetism, National Academy of Sciences of Ukraine, 36b Vernadskogo Boulevard, 03142 Kyiv, Ukraine
}%
\author{Pavlo Bondarenko}%
\affiliation{%
Institute of Magnetism, National Academy of Sciences of Ukraine, 36b Vernadskogo Boulevard, 03142 Kyiv, Ukraine
}%
\author{Elena V. Tartakovskaya}%
\affiliation{%
Institute of Spintronics and Quantum Information, Faculty of Physics, Adam Mickiewicz University, Pozna\'{n}, Uniwersytetu Pozna\'{n}skiego 2, 61-614 Pozna\'{n}, Poland 
}%
\affiliation{%
Institute of Magnetism, National Academy of Sciences of Ukraine, 36b Vernadskogo Boulevard, 03142 Kyiv, Ukraine
}%
\author{Maciej~Krawczyk}%
\affiliation{%
Institute of Spintronics and Quantum Information, Faculty of Physics, Adam Mickiewicz University, Pozna\'{n}, Uniwersytetu Pozna\'{n}skiego 2, 61-614 Pozna\'{n}, Poland 
}%

\date{\today}

\begin{abstract}
The spin-wave dynamics in the thin CoFeB film in Damon-Eshbach geometry are studied in three cases of boundary conditions---free boundary conditions, symmetrical surface anisotropy, and one-sided surface anisotropy. The analytical model created by Wolfram and De Wames was extended to include perpendicular surface anisotropy in boundary conditions. Its comparison with numerical simulations demonstrate perfect agreement between the approaches. The analysis of the dispersion relation indicates that the presence of surface anisotropy increases the avoided crossing size between Damon-Eshbach mode and perpendicular standing modes. Additionally, asymmetrical one-sided surface anisotropy induces nonreciprocity in the dispersion relation. In-depth analysis of the avoided crossing size is conducted for systems with different boundary conditions, different thicknesses, surface anisotropy constant values, and external magnetic fields. It shows the significant role of the strength of surface localization of Damon-Eshbach mode and the symmetry of perpendicular standing modes in the avoided crossing broadening. Interestingly, for specific set of parameters the interaction between the particular modes can be suppressed, resulting in a mode crossing. Such a crossing, which occurs only on one side of the dispersion relation in a one-sided surface anisotropy system, can be utilized in nonreciprocal devices.
\end{abstract}

\maketitle



\section{Introduction}

In recent years, spin waves (SWs), which are collective, harmonic oscillations of spins that propagate within magnetic materials, have received increased attention due to their potential to transport and process information with the reduction of Joule heating and energy dissipation \cite{Chu15}. One of the interesting properties of propagating SWs in thin magnetic films in Damon-Eshbach (DE) geometry \cite{Dam61} is the hybridization between the fundamental SW mode and perpendicular standing SW (PSSW) modes \cite{Gru82,Zha21,Gla16,Tac19,Wang22,Dre21,Song21}. This may result in the formation of avoided crossings (ACs), which can be a crucial physical characteristic for the development of magnonic devices such as filters and phase shifters. However, the control of the dynamic magnetic properties is a fundamental problem for the implementation of these devices.

It has been demonstrated that surface anisotropy significantly impacts the dispersion relation and the AC size between propagating SW mode and PSSW modes \cite{Van21}. Another studies have shown that surface anisotropy can be controlled by the voltage applied across the ferromagnetic-metal/insulator heterostructures due to the charge accumulation at the interface \cite{Rana19,Wang17,Rana18} or across insulator/ferromagnet/insulator multilayer due to the dielectric polarization influence on the interface \cite{Ibr16}. Therefore, it can be concluded that hybridization between fundamental SW mode and higher-order PSSW modes could be controlled by electric field. However, there has been no systematic study on the influence of surface anisotropy on the hybridization between SW modes in the ferromagnetic film. 

In general, there are two alternative approaches which can be used for the analytical evaluation of dipole-exchange SW spectrum including interaction between fundamental SW mode and PSSW modes. One approach, proposed by Wolfram and De Wames \cite{Wam70,Wol70}, involves solving a sixth-order differential equation derived from Maxwell's equations along with equations of the magnetization motion. The extension of Damon and Eshbach's theory for pure dipolar SWs by including exchange interactions provides evidence that, as a result of exchange, the surface and bulk modes mix. This theoretical approach was used for explanation of the first experiments on magnon branch repulsion in thin ferromagnetic films with in-plane magnetization \cite{Cam80,Kab84} and in thin single-crystal disks of yttrium-iron garnet \cite{Hen72}. Much later, researchers applied the same method to characterize SWs in infinitely long cylindrical wires with magnetization along the wire \cite{Ari01, Ryc18}.

However, it turned out that the Wolfram and De Wames approach is not suitable for a broad range of sample geometries and magnetic moment directions. In fact, its effectiveness is limited to cases of unbroken symmetry in infinite films, as well as in infinite wires with a magnetic moment along the wire axis, as previously noted. For more general cases, Kalinikos and Slavin proposed an alternative approach for mixed exchange boundary conditions in thin films and the arbitrary direction of external magnetic field and magnetic moment relative to the film plane \cite{Kal86,Kal90}. The first step of this method is to solve Maxwell's equations separately in the magnetostatic approximation \cite{Akh68}. Then, the dynamical scalar potential obtained in the form of the tensorial magnetostatic Green’s functions \cite{Gus11} is inserted into the equations of motion for the magnetic moment (linearized Landau-Lifshitz equations), and the resulting integro-differential equation is solved through perturbation theory. This method has resolved most theoretical issues of spin dynamics in laterally confined magnetic elements under different magnetic field configurations. It has been previously applied to describe SW dynamics in isolated magnetic stripes \cite{Gus02} as well as rectangular \cite{Gub04,Bay05}, cylindrical \cite{Gus00}, and triangular \cite{Kha19} magnetic dots. A notable benefit of the Kalinikos and Slavin method is that it utilizes a simple analytical formula to achieve good agreement between theory and experiment for thin, circular nanoelements with perpendicularly-magnetized states, such as rings \cite{Zhou21} and dots \cite{Kak04}. In more complex cases with broken cylindrical symmetry, it is necessary to consider a greater number of perturbation theory terms (i.e., the interaction of SW modes) \cite{Tar16,Tar05}. However, the applicability of this theory to any case of nanostructures and geometry of applied fields is not in question. The method of Wolfram and De Wames turned out to be somewhat forgotten, which forced Harms and Duine \cite{Har22} to "rediscover" this ansatz since in some cases it provides a more direct path to the result.

A comprehensive review of the two mentioned approaches with an analysis of their applicability for various cases of the direction of the external field and magnetization in a ferromagnetic film is given by Arias \cite{Ari16}. The potential drawbacks of the Kalinikos-Slavin method were identified, including possible inaccuracies of the results obtained in the region of hybridization of SW modes, as well as the complexity of describing the interaction of surface and bulk modes. The theoretical approach proposed in \cite{Ari16} is based on the method developed by Wolfram and De Wames and provides strict solutions to the problem. It is important to note that the hybridization of SWs was only examined in the case of mixed symmetrical boundary conditions.

In this paper, we conduct a systematical analysis of the impact of surface anisotropy on the SW hybridization, which was presented in \cite{Van21}. We are confronted with a choice between the two methods described above for calculating the dynamics of SWs should be chosen. Following the conclusions of Arias \cite{Ari16}, the Wolfram and De Wames method not only leads to the goal more efficiently in this case, despite the asymmetry of the boundary conditions, but also provides a rigorous solution. This is in contrast to the Kalinikos-Slavin perturbation theory which requires a significant number of iterations and provides only an approximate solution. Therefore, we compared the dispersion relations of SWs in an DE geometry using symmetrical and asymmetrical boundary conditions via the extended Wolfram and De Wames approach. The results of analytical calculations perfectly matched the numerical simulations on the example of CoFeB thin film. We provide an in-detail analysis of the dispersion relations, SW mode profiles, and the effect of material parameters on the SW coupling in the frame of AC size.


\section{Methods}

\subsection{Investigated system}

\begin{figure}[t]
    \includegraphics{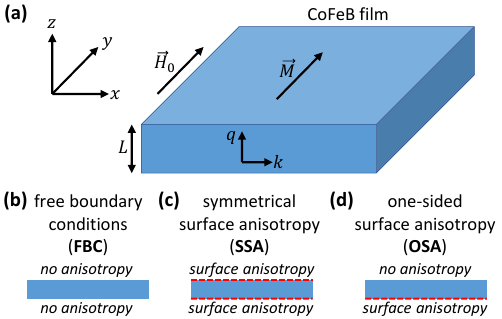}
    \caption{(a) A general schematic of the system and coordinate system. (b-d) Schematics of the boundary conditions investigated in the manuscript: (b) free boundary conditions, (c) symmetrical surface anisotropy, and (d) one-sided surface anisotropy.}
    \label{fig:figure1}
\end{figure}

The system under investigation is presented in Fig.~\ref{fig:figure1}(a). It is a thin CoFeB film of thickness $L$ magnetized in-plane in $y$-direction by the external magnetic field $H_0$. We consider the DE geometry, i.e., the SWs propagating along the $x$-direction, perpendicular to the external field $H_0$. The $z$-axis corresponds to the direction perpendicular to the film plane, where the surfaces of the film are located at $z=\pm L/2$. The following parameters were used for CoFeB: magnetization saturation $M_\mathrm{S} = \SI{1335}{\kilo\ampere/\meter}$, exchange stiffness $A_\mathrm{ex} = \SI{15}{\pico\joule/\meter}$, and gyromagnetic ratio $\gamma = \SI{30}{\giga\hertz/\tesla}$. In this study, we consider three cases of boundary conditions: free boundary conditions (FBC) where the surface anisotropy is absent in the system [Fig.~\ref{fig:figure1}(b)]; symmetrical surface anisotropy (SSA), i.e., the surface anisotropy of equal strength is present on both boundaries of the film [Fig.~\ref{fig:figure1}(c)]; one-sided surface anisotropy (OSA) where the bottom surface has non-zero surface anisotropy while the top surface is described with FBC [Fig.~\ref{fig:figure1}(d)].

\subsection{Analytical model}

We use the approach proposed by Wolfram and De Wames \cite{Wam70,Wol70} to calculate the dispersion relation in DE geometry in dipole-exchange regime and extend it to include the perpendicular surface anisotropy introduced by Rado and Weertman \cite{Rad59}.

The magnetic free energy of the system can be presented as
\begin{equation}\label{eq:energy}
    F = \int \left( -\mu_0 \mathbf{H}_0 \cdot \mathbf{M} + \frac{A_\mathrm{ex}}{M_\mathrm{S}^2} \left( \nabla \mathbf{M} \right)^2 - \frac{1}{2} \mu_0 \mathbf{H}_\mathrm{d} \cdot \mathbf{M} \right) \mathrm{d}V,
\end{equation}
where there are three terms in the integral---the first term represents the Zeeman energy, the second term represents the exchange energy, and the third term represents the magnetostatic energy, $\mathbf{M}$ is the magnetization vector, $\mu_0$ is the vacuum permeability, $\mathbf{H}_\mathrm{d}$ is the demagnetizing field.

The dynamics of the magnetic system are described with Landau-Lifshitz equation
\begin{equation}\label{eq:llg}
    \frac{\partial \mathbf{M}}{\partial t} = -|\gamma| \mu_0 \mathbf{M} \times \mathbf{H}_\mathrm{eff},
\end{equation}
where $\mathbf{H}_\mathrm{eff} = -\frac{1}{\mu_0} \frac{\delta F}{\delta \mathbf{M}}$ is the effective magnetic field.

The demagnetizing field $\mathbf{H}_\mathrm{d}$ is derived from the Maxwell equations in magnetostatic approximation:
\begin{equation}\label{eq:maxwell}
    \nabla \times \mathbf{H}_\mathrm{d} = 0, \,\,\,\,\, \nabla \cdot \mathbf{B} = 0,
\end{equation}
where  $\mathbf{B} = \mu_0 (\mathbf{H}_\mathrm{d} + \mathbf{M})$ is the magnetic induction. Equation~(\ref{eq:maxwell}) enables the introduction of magnetic scalar potential $\varphi$, which satisfies the formula $\mathbf{H}_\mathrm{d} = -\nabla \varphi$. As a result, the magnetostatic Maxwell equations are replaced with a single equation for the magnetic scalar potential
\begin{equation}
    \Delta \varphi = \nabla \cdot \mathbf{M}.
\end{equation}

Thanks to the uniform magnetization, the Landau-Lifshitz equation can be easily linearized. We assume that the static $y$-component of the magnetization remains constant and is equal to the saturation magnetization $M_\mathrm{S}$, while the dynamic component $\mathbf{m}=(m_x,m_z)$, which is much smaller than the static component $M_y$ ($|\mathbf{m}|\ll M_\mathrm{S}$), precesses in the $xz$-plane. Therefore, $\mathbf{M}(x,y,z,t)=M_\mathrm{S} \hat{y} + \mathbf{m}(x,z)e^{i\omega t}$, where $\omega = 2 \pi f$ is the angular frequency and $f$ is the frequency.

After linearization, the SW dynamics are described with a set of three coupled equations:
\begin{equation}\label{eq:mx}
    i\omega m_x = \gamma \mu_0 \left(H_0-\frac{2 A_\mathrm{ex}}{\mu_0 M_\mathrm{S}}\Delta\right) m_z + M_\mathrm{S} \partial_z \varphi,
\end{equation}
\begin{equation}\label{eq:mz}
    -i\omega m_z = \gamma \mu_0 \left(H_0-\frac{2 A_\mathrm{ex}}{\mu_0 M_\mathrm{S}}\Delta\right) m_x + M_\mathrm{S} \partial_x \varphi,
\end{equation}
\begin{equation}\label{eq:pot}
    \Delta \varphi - \partial_x m_x - \partial_z m_z = 0.
\end{equation}

The solutions to Eqs.~(\ref{eq:mx})-(\ref{eq:pot}) take the form of plane waves. Two wave vectors can be defined due to the system's symmetry: in-plane wave vector $k$ (in the $x$-direction) and out-of-plane wave vector $q$ (in the $z$-direction), as shown in Fig.~\ref{fig:figure1}(a). As a result, we have $(m_x,m_z,\varphi) \propto (m_{x0},m_{z0},\varphi_0 ) e^{ikx} e^{iqz}$. The system in the $x$-direction is infinite, therefore the wave vector $k$ can only have real values for the solution to be physical. On the other hand, the wave vector $q$ may take on complex values. For simplicity, we introduce the following dimensionless parameters: $\Omega = \frac{\omega}{\gamma \mu_0 M_\mathrm{S}}$, $\theta = \Omega_H + \lambda^2 (k^2+q^2)$, $\Omega_H = \frac{H_0}{M_\mathrm{S}}$, and $\lambda^2 = \frac{2 A_\mathrm{ex}}{\mu_0 M_\mathrm{S}^2}$. After substituting the plane-wave solution into Eqs.~(\ref{eq:mx})-(\ref{eq:pot}) and expressing them in the matrix form, we obtain
\begin{equation}\label{eq:matrix}
    \begin{pmatrix}
        i\Omega & \theta & iq \\
        \theta & -i\Omega & ik \\
        ik & iq & k^2+q^2
    \end{pmatrix}
    \begin{pmatrix}
        m_{x0} \\
        m_{z0} \\
        \varphi_0
    \end{pmatrix}=0.
\end{equation}
The condition that the determinant of the 3x3 matrix in Eq.~(\ref{eq:matrix}) is equal to zero leads to the following formula:
\begin{equation}\label{eq:kqomega}
    (k^2+q^2) (\Omega^2 - \theta^2 - \theta) = 0.
\end{equation}
As $\theta = \theta(q^2)$, Eq.~(\ref{eq:kqomega}) is a third-degree function with respect to $q^2$. Two roots, $q=\pm ik$, are obtained by setting the first bracket to zero whereas four roots, $q=\pm q_1$ and $q=\pm iq_2$ where $q_1,q_2\in \mathbb{R}$, are obtained by setting the second bracket to zero. From the zeroing of the second bracket in Eq.~(\ref{eq:kqomega}), we can also derive the formula for the dimensionless frequency
\begin{equation}\label{eq:omega}
    \Omega = \sqrt{\theta (\theta+1)}.
\end{equation}
Let $\theta (q=q_1) = \theta_1$ and $\theta (q=q_2) = \theta_2$. Since $q_1$ and $q_2$ correspond to the same frequency, $\Omega = \sqrt{\theta_1 (\theta_1+1)} = \sqrt{\theta_2 (\theta_2+1)}$, and therefore, $\theta_2 = -(\theta_1+1)$. From this formula we can obtain the connection between wave vectors $k$, $q_1$, and $q_2$, which is the following:
\begin{equation}\label{eq:q2}
    q_2 = \pm \sqrt{2k^2 + q_1^2 + \frac{2\Omega_H+1}{\lambda^2}}.
\end{equation}

We can interpret the solutions obtained for the out-of-plane wave vector $q$ as follows. Since our solution is a plane wave, wave vector $q_1$ will give a volume contribution of the sinusoidal character to the mode profile while wave vectors $k$ and $q_2$ denote exponentially-decaying modes localized on the surfaces. Since the wave vector $k$ represents also the propagating in-plane wave vector, this solution has a character of a DE mode. Next, knowing that $\Omega_H\geq 0$, we can derive from Eq.~(\ref{eq:q2}) that $|q_2|\geq 1/\lambda$, indicating that $q_2$ has a character of a surface exchange mode.

\begin{widetext}

The solution of Eq.~(\ref{eq:matrix}) can be represented by a vector
\begin{equation}\label{eq:m0}
    \begin{pmatrix}
        m_{x0} \\
        m_{z0} \\
        \varphi_0
    \end{pmatrix}
    =
    \begin{pmatrix}
        ik\theta - q\Omega \\
        iq\theta + k\Omega \\
        \Omega^2 - \theta^2
    \end{pmatrix} C,
\end{equation}
where $C$ is an arbitrary constant. The general solution for the full vector $(m_x,m_z,\varphi)$ is a superposition of six terms, one for each solution of the wave vector $q$
\begin{equation}\label{eq:profile}
    \begin{pmatrix}
        m_{x} \\
        m_{z} \\
        \varphi
    \end{pmatrix}
    =
    \left[
    \begin{pmatrix}
        X_1 \\
        Z_1 \\
        F_1
    \end{pmatrix} C_1 e^{iq_1z}
    +
    \begin{pmatrix}
        X_2 \\
        Z_2 \\
        F_2
    \end{pmatrix} C_2 e^{-iq_1z}
    +
    \begin{pmatrix}
        X_3 \\
        Z_3 \\
        F_3
    \end{pmatrix} C_3 e^{kz}
    +
    \begin{pmatrix}
        X_4 \\
        Z_4 \\
        F_4
    \end{pmatrix} C_4 e^{-kz}
    +
    \begin{pmatrix}
        X_5\\
        Z_5 \\
        F_5
    \end{pmatrix} C_5 e^{q_2z}
    +
    \begin{pmatrix}
        X_6 \\
        Z_6 \\
        F_6
    \end{pmatrix} C_6 e^{-q_2z}
    \right] e^{ikx}
\end{equation}
where $X_1 = ik\theta_1 - q_1\Omega$, $X_2 = ik\theta_1 + q_1\Omega$, $X_3 = X_4 = ik$, $X_5 = ik\theta_2 - iq_2\Omega$, $X_6 = ik\theta_2 + iq_2\Omega$, $Z_1 = k\Omega + iq_1\theta_1$, $Z_2 = k\Omega - iq_1\theta_1$, $Z_3 = -k$, $Z_4 = k$, $Z_5 = k\Omega - q_2\theta_2$, $Z_6 = k\Omega + q_2\theta_2$, $F_1 = F_2 = \Omega^2 - \theta_1^2$, $F_3 = -(\Omega + \Omega_H)$, $F_4 = \Omega - \Omega_H$, $F_5 = F_6 = \Omega^2 - \theta_2^2$, as it follows from Eq.~(\ref{eq:m0}).
\end{widetext}

As the system under consideration is an infinite film, boundary conditions must be applied on top and bottom surfaces. Our goal was to extend the model derived by Wolfram and De Wames to include the presence of the perpendicular surface anisotropy. It requires the extension of exchange boundary condition by adding the term depending on the surface anisotropy \cite{Rad59}
\begin{equation}\label{eq:bc}
  \left\{
  \begin{aligned}
    \partial_z m_x       & = 0 |_{z=\pm L/2}\\
    \partial_z m_z \mp \sigma_\mathrm{t(b)}m_z  & = 0 |_{z=\pm L/2}
  \end{aligned}
  \right.
\end{equation}
where $\sigma_\mathrm{t(b)} =K_\mathrm{s}^\mathrm{t(b)} /A_\mathrm{ex}$ and $K_\mathrm{s}^\mathrm{t(b)}$ is surface anisotropy constant for the top (bottom) surface.

Since the equation for the magnetic scalar potential [Eq.~(\ref{eq:pot})] outside of the film gives $\Delta \varphi_\mathrm{out} = 0$ and, subsequently, $-\varphi_0 (k^2+q^2) e^{ikx} e^{iqz} = 0$, the asymptotic solutions outside the film for the magnetic scalar potential are given by expression
\begin{equation}\label{eq:bcpot}
  \varphi_\mathrm{out} = 
  \begin{cases}
      C_7 e^{ikx} e^{-|k|z} & \text{for } z \geq L/2,\\
      C_8 e^{ikx} e^{|k|z}  & \text{for } z \leq -L/2.
  \end{cases}
\end{equation} 
As the tangential components of the demagnetizing field $\mathbf{H}_\mathrm{d}$ are continuous across the surfaces of the film, the magnetic scalar potential must also be continuous. Additionally, the normal component of $\mathbf{B}$ must also be continuous. Therefore, this results in the effective magnetostatic boundary conditions:
\begin{equation}\label{eq:varphiout}
    \varphi = \varphi_\mathrm{out},
\end{equation} 
\begin{equation}\label{eq:Bz}
    B_z = B_z^\mathrm{out},
\end{equation} 
where $\varphi$ and $B_z$ are magnetic scalar potential and magnetic induction in the magnetic material, and $\varphi_\mathrm{out}$ and $B_z^\mathrm{out}$ -- out of the magnetic material, respectively. Then, Eq.~(\ref{eq:Bz}) can be rewritten in terms of scalar potential as
\begin{equation}\label{eq:Bzfull}
    \partial_z \varphi - m_z = \partial_z \varphi_\mathrm{out}.
\end{equation}
The complete set of boundary conditions in Eqs.~(\ref{eq:bc}),~(\ref{eq:varphiout}), and (\ref{eq:Bzfull}) evaluated for the SW modes in Eq.~(\ref{eq:profile}) leads to the following degeneracy matrix $\boldsymbol{A}$:

\begin{widetext}
\begin{equation}\label{eq:thisridiculousmatrix}
\begin{split}
    \boldsymbol{A} = 
    \left(
    \begin{matrix}
        iq_1 X_1 e^{iq_1\frac{L}{2}} & -iq_1 X_2 e^{-iq_1\frac{L}{2}} & k X_3 e^{k\frac{L}{2}} \\
        iq_1 X_1 e^{-iq_1\frac{L}{2}} & -iq_1 X_2 e^{iq_1\frac{L}{2}} & k X_3 e^{-k\frac{L}{2}} \\
        (iq_1 - \sigma_\mathrm{t}) Z_1 e^{iq_1\frac{L}{2}} & (-iq_1 - \sigma_\mathrm{t}) Z_2 e^{-iq_1\frac{L}{2}} & (k - \sigma_\mathrm{t}) Z_3 e^{k\frac{L}{2}} \\
        (iq_1 + \sigma_\mathrm{b}) Z_1 e^{-iq_1\frac{L}{2}} & (-iq_1 + \sigma_\mathrm{b}) Z_2 e^{iq_1\frac{L}{2}} & (k + \sigma_\mathrm{b}) Z_3 e^{-k\frac{L}{2}} \\
        [(iq_1 + |k|)F_1 - Z_1] e^{iq_1\frac{L}{2}} & [(-iq_1 + |k|)F_2 - Z_2] e^{-iq_1\frac{L}{2}} & [(k + |k|)F_3 - Z_3] e^{k\frac{L}{2}} \\
        [(iq_1 - |k|)F_1 - Z_1] e^{-iq_1\frac{L}{2}} & [(-iq_1 - |k|)F_2 - Z_2] e^{iq_1\frac{L}{2}} & [(k - |k|)F_3 - Z_3] e^{-k\frac{L}{2}}
    \end{matrix}
    \right|
    \\
    \left|
    \begin{matrix}
        -k X_4 e^{-k\frac{L}{2}} & q_2 X_5 e^{q_2\frac{L}{2}} & -q_2 X_6 e^{-q_2\frac{L}{2}} \\
        -k X_4 e^{k\frac{L}{2}} & q_2 X_5 e^{-q_2\frac{L}{2}} & -q_2 X_6 e^{q_2\frac{L}{2}} \\
        (-k - \sigma_\mathrm{t}) Z_4 e^{-k\frac{L}{2}} & (q_2 - \sigma_\mathrm{t}) Z_5 e^{q_2\frac{L}{2}} & (-q_2 - \sigma_\mathrm{t}) Z_6 e^{-q_2\frac{L}{2}} \\
        (-k + \sigma_\mathrm{b}) Z_4 e^{k\frac{L}{2}} & (q_2 + \sigma_\mathrm{b}) Z_5 e^{-q_2\frac{L}{2}} & (-q_2 + \sigma_\mathrm{b}) Z_6 e^{q_2\frac{L}{2}} \\
        [(-k + |k|)F_4 - Z_4] e^{-k\frac{L}{2}} & [(q_2 + |k|)F_5 - Z_5] e^{q_2\frac{L}{2}} & [(-q_2 + |k|)F_6 - Z_6] e^{-q_2\frac{L}{2}} \\
        [(-k - |k|)F_4 - Z_4] e^{k\frac{L}{2}} & [(q_2 - |k|)F_5 - Z_5] e^{-q_2\frac{L}{2}} & [(-q_2 - |k|)F_6 - Z_6] e^{q_2\frac{L}{2}} 
    \end{matrix}
    \right).
\end{split}
\end{equation}
\end{widetext}

The condition $\det{\boldsymbol{A}} = 0$ allows obtaining the solutions of wave vector $q$ and, subsequently, the resonance frequencies as a function of wave vector $k$. The eigenvectors of matrix $\boldsymbol{A}$ provide the coefficients $C_i$ in Eq.~(\ref{eq:profile}).

Compared to the approach suggested by Kalinikos et al. \cite{Kal90}, the solution mentioned above is precise within the examined geometry. Calculating multiple integrals for components of a demagnetizing tensor and expanding dynamical magnetization components into a series is not required to obtain coupled modes, which simplifies analytical calculations and significantly reduces computation time.

\subsection{Numerical simulations}

The Landau-Lifshitz equation in the linear approximation [Eqs.~(\ref{eq:mx}),(\ref{eq:mz})] and the magnetostatic Maxwell equation-based formula for the magnetic scalar potential [Eq.~(\ref{eq:pot})] along with the boundary conditions for perpendicular surface anisotropy [Eq.~(\ref{eq:bc})] and magnetostatic potential [Eq.~(\ref{eq:bcpot})] were solved numerically using finite-element method simulations in COMSOL Multiphysics \cite{Van21}. The problem was solved in 1D geometry with reduced $x$- and $y$-directions. Eqs.~(\ref{eq:mx})-(\ref{eq:pot}) were modified accordingly to introduce the terms coming from the implementation of plane-wave solution representing the propagation of SWs in $x$-direction $(m_x,m_z,\varphi) = (m_{x0},m_{z0},\varphi_0) e^{ikx}$. The dispersion relations were calculated using eigenfrequency study.


\section{Results and discussion}

\subsection{Dispersion relation analysis}

\begin{figure*}[t]
    \includegraphics{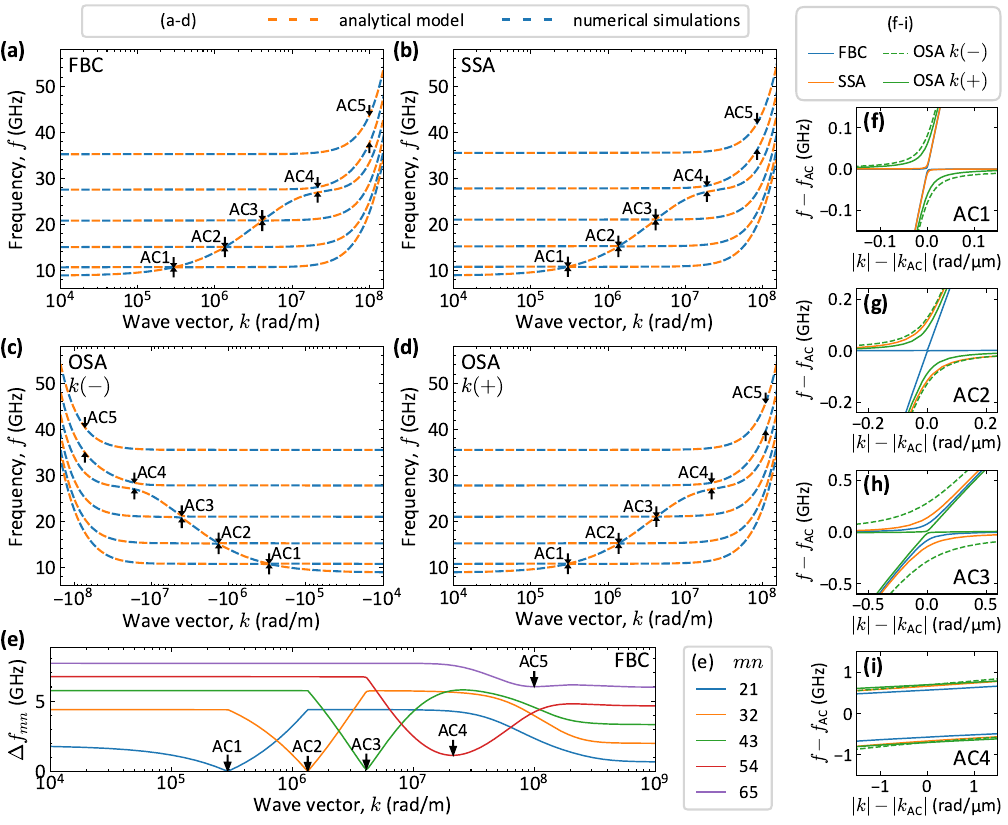}
    \caption{(a-d) Dispersion relations of six lowest modes of a 100 nm-thick CoFeB film with (a) FBC, (b) SSA with $K_\mathrm{s}^\mathrm{t}=K_\mathrm{s}^\mathrm{b}=\SI{-700}{\micro \joule /\meter^2}$, and (c,d) OSA with $K_\mathrm{s}^\mathrm{t}=0$ and $K_\mathrm{s}^\mathrm{b}=\SI{-1500}{\micro \joule /\meter^2}$ for (c) negative and (d) positive wave vectors in the external magnetic field $\mu_0 H_0 = \SI{50}{\milli \tesla}$. The plots present the comparison between the analytical model (orange lines) and numerical simulations (blue lines). Avoided crossings (ACs) are marked with labels. (e) The frequency difference between neighboring modes in FBC system. In plots (a-e) wave vector $k$ on the $x$-axis is presented in the logarithmic scale. (f-i) Close-up on the ACs: (f) AC1, (g) AC2, (h) AC3, and (i) AC4. The plot axis are showing the wave vector and frequency values relative to the AC position calculated from Eqs.~(\ref{eq:dfn})~and~(\ref{eq:dfn2}), respectively. Plots present numerical simulations results only which are in agreement with analytical results.}
    \label{fig:dispn}
\end{figure*}

First, we study the effect of the surface anisotropy on the dispersion relation. We chose the thickness of the CoFeB film $L = \SI{100}{\nano\meter}$ and external magnetic field $\mu_0 H_0 = \SI{50}{\milli\tesla}$. We show the dispersion relation of the six lowest modes for three cases---free boundary conditions (FBC), i.e., $K_\mathrm{s}^\mathrm{t}=K_\mathrm{s}^\mathrm{b}=0$ [Fig.~\ref{fig:dispn}(a)]; symmetrical surface anisotropy (SSA) with $K_\mathrm{s}^\mathrm{t}=K_\mathrm{s}^\mathrm{b}=\SI{-700}{\micro \joule /\meter^2}$ [Fig.~\ref{fig:dispn}(b)]; one-sided surface anisotropy (OSA) with $K_\mathrm{s}^\mathrm{t}=0$ and $K_\mathrm{s}^\mathrm{b}=\SI{-1500}{\micro \joule /\meter^2}$ separately for negative [Fig.~\ref{fig:dispn}(c)] and positive [Fig.~\ref{fig:dispn}(d)] wave vector $k$. Values of surface anisotropy are comparable to the values presented in literature \cite{Joh96}.

The dispersion relation calculated with the analytical model is shown as a dashed orange line while the numerical simulation results are shown with dashed blue line. Figs.~\ref{fig:dispn}(a-d) demonstrate the perfect agreement between these two methods, yielding identical outcomes. The nature of dispersions is characteristic of the system in DE geometry. Each plot consists of one branch with a significant slope in the center of the investigated range of wave vector $k$, displaying a DE surface mode character, and the remaining five are flat branches representing PSSW modes. All the modes start to increase significantly in frequency at about $10^7$ rad/m as a result of the increasing contribution of the exchange interaction to the SW energy. Positions of PSSW modes at $k\approx 0$ are determined by the wave vector $q_1 \approx n\pi /L$ ($n=1,2,3...$ is the PSSW mode number). In the presence of negative surface anisotropy, the value of $q_1 > n\pi /L$ for corresponding PSSW modes (the reverse happens for positive surface anisotropy). The increase of the frequency of the DE mode correlates with the increase of its wave vector $q_1$ with the increase of $k$. However, $q_1$ begins to decrease at some point, leading to $q_1\approx n\pi/L$ for very large wave vectors $k$. Similarly as for the case of $k \approx 0$, for very large $k$ in the presence of negative surface anisotropy, $q_1 > n\pi /L$ (the reverse happens for positive surface anisotropy). Detailed explanation of the correlation between wave vectors $k$ and $q_1$ is provided in Appendix~\ref{app1}. The DE mode increases in frequency and intersects with the three lowest PSSW modes, leading to the emergence of ACs. These ACs are labeled in Figs.~\ref{fig:dispn}(a-d) with the abbreviation AC and a number indicating their sequence, beginning with the lowest.

The discussion of ACs requires a precise definition of where AC occurs. Neglecting the atomic distance limit, the theory provides infinite number of SW modes. Though it is hypothetically possible for AC to be present between all modes, it is apparent that the number of ACs is not infinite for the finite-thickness film. To denote the presence of AC, we establish two distinct criteria. The first is the local minimum criterion. If the function that represents the frequency difference between the neighboring modes 
\begin{equation}\label{eq:dfn}
    \Delta f_{mn} = f_{m}(k)-f_n(k)
\end{equation}
(where $m,n$ is a mode number) has a local minimum $\Delta f_{\mathrm{AC}n}$, this minimum represents an AC (or simply crossing if $\Delta f_\mathrm{AC} = 0$). In this way, we can define an AC for any boundary conditions and it allows multiple ACs if multiple local minima exist. The second is a frequency limit criterion. It could be clearly defined only for FBC. It says that an AC is present between the DE mode and $n$-th PSSW mode if $f_{k\to\infty}^\mathrm{DE} > f_{k=0}^n$ in case where $f_{k\to\infty}^\mathrm{DE}$ is calculated for $A_\mathrm{ex}=0$ \cite{Dam61}, i.e. 
\begin{equation}\label{eq:flcDE}
    f_{k\to\infty}^\mathrm{DE} = \frac{\mu_0 \gamma}{2\pi} \left( H_0+\frac{M_\mathrm{S}}{2} \right)
\end{equation}
and \cite{Gurevich}
\begin{eqnarray}\label{eq:flcPSSW}
    f_{k=0}^n = \frac{\mu_0 \gamma}{2\pi}  \left(\left(H_0+\frac{2A_\mathrm{ex}}{\mu_0 M_\mathrm{S}} \left(\frac{n\pi}{L}\right)^2\right)\right.\times\nonumber\\   \left.\times\left(H_0+M_\mathrm{S}+\frac{2A_\mathrm{ex}}{\mu_0 M_\mathrm{S}} \left(\frac{n\pi}{L}\right)^2\right)\right)^{1/2}
\end{eqnarray}
This criterion is valid under the assumption that the contribution of the exchange interaction to the $k$ dependence of the frequency of DE mode and PSSW modes is identical. The AC position is determined by the minimum of Eq.~(\ref{eq:dfn}). It means that the choice of criterion does not influence the value of the AC size. In this paper, we present the results based on the local minimum criterion because of its broader definition. However, we will also mention the frequency limit criterion and its impact on the results.

To address AC occurrence accurately, the frequency difference between neighboring modes is presented as a function of wave vector $k$ in Fig.~\ref{fig:dispn}(e) for the case of FBC, for which the dispersion relation is shown in Fig.~\ref{fig:dispn}(a). In the range of small and large wave vectors, the distance between the modes is almost constant. The discrepancy between these ranges is due to the fact that in the limit of small wave vectors, the dispersion relation of the modes can be described by Eq.~(\ref{eq:flcPSSW}) \cite{Gurevich}, while in the large wave vector limit with the function 
\begin{equation}\label{eq:largek}
    f_n = \frac{\mu_0 \gamma}{2\pi} \left(H_0+M_\mathrm{S}+\frac{2A_\mathrm{ex}}{\mu_0 M_\mathrm{S}}k^2+\frac{2A_\mathrm{ex}}{\mu_0 M_\mathrm{S}} \left(\frac{n\pi}{L}\right)^2\right).
\end{equation}

\begin{table*}[t]
    \centering
    \caption{AC size of AC1-AC5 for FBC, SSA, and OSA systems, which dispersion relations are shown in Figs.~\ref{fig:dispn}(a-d).}
    \begin{ruledtabular}
    \begin{tabular}{cddddd}
        System     & \multicolumn{1}{c}{\textrm{AC1 (MHz)}} & \multicolumn{1}{c}{\textrm{AC2 (MHz)}} & \multicolumn{1}{c}{\textrm{AC3 (MHz)}} & \multicolumn{1}{c}{\textrm{AC4 (MHz)}} & \multicolumn{1}{c}{\textrm{AC5 (MHz)}} \\ \hline
        FBC        & 11.14 &  6.04 & 158.8 & 1137.2 & 6022.4 \\
        SSA        & 21.03 & 231.5 & 280.6 & 1327.7 & 5781.5 \\
        OSA ($k-$) & 162.8 & 254.8 & 565.6 & 1389.2 & 5474.4 \\
        OSA ($k+$) & 122.8 & 175.7 & 24.67 & 1396.0 & 6119.0 \\
    \end{tabular}
    \end{ruledtabular}
    \label{tab:tab}
\end{table*}

In the mid-range, each curve shown in Fig.~\ref{fig:dispn}(e) has a local minimum corresponding to the AC, which is labeled and marked with an arrow. The first three ACs are relatively small, not exceeding a size of 200 MHz.  The AC4, represented by a deep minimum, has a size of 1.14 GHz. On the other hand, AC5 has a very shallow minimum with a size of 6.02 GHz. Interestingly, it is not the global minimum, as according to Eq.~(\ref{eq:largek}) the distance between the modes can reach 5.99 GHz, which value is in agreement with the analytical model. However, according to the local minimum criterion, it is considered to be an AC. In case of the frequency limit criterion, only the first three minima can be identified as ACs. The AC4 does not meet this criterion as $f_{k=0}^{n=4} = \SI{27.55}{\giga\hertz}$ exceeds $f_{k\to\infty}^\mathrm{DE} = \SI{26.66}{\giga\hertz}$ slightly.

After presenting the similarities between the systems, it is time to highlight the differences. Firstly, the symmetry of the system, specifically the boundary conditions, leads to the symmetry of the dispersion relation with respect to wave vector. Therefore, the FBC and SSA systems have symmetrical dispersions since $K_\mathrm{s}^\mathrm{t}=K_\mathrm{s}^\mathrm{b}$. In contrast, the OSA system has different surface anisotropy constants on the top and bottom surfaces, resulting in a frequency difference between negative and positive wave vectors. Additionally, the presence of the negative surface anisotropy causes a slight increase in the frequency of all modes. Comparing the results in Figs.~\ref{fig:dispn}(a)~and~(b), for $K_\mathrm{s} = \SI{-700}{\micro \joule /\meter^2}$ the increase does not pass 1 GHz. Conversely, for a positive surface anisotropy, a decrease in frequency would be noted.

The most significant difference between the systems lies in the size of the ACs. Close-up plots are shown in Fig.~\ref{fig:dispn}(f-i) for AC1-AC4, respectively. They show the dispersion relation for the values of wave vector and frequency relative to the AC location ($k_\mathrm{AC}$,$f_\mathrm{AC}$), which is defined separately for each AC in the following way –- $k_{\mathrm{AC}_n}$ represents the wave vector of the local minimum of Eq.~(\ref{eq:dfn}), while $f_{\mathrm{AC}_n}$
\begin{equation}\label{eq:dfn2}
    f_{\mathrm{AC}_n} = \frac{f_n(k_{\mathrm{AC}_n})+f_{n+1}(k_{\mathrm{AC}_n})}{2} .
\end{equation}
The values of the AC size for each AC type can be found in Table~\ref{tab:tab}. AC1 [Fig.~\ref{fig:dispn}(f)] exhibits a negligible size for the FBC and SSA systems, but a more significant size of 162.8 MHz for negative and 122.8 MHz for positive wave vectors in the OSA system. As the dispersion relations for FBC and SSA are symmetrical, the AC sizes are always equal for both negative and positive wave vectors. The size of AC2 [Fig.~\ref{fig:dispn}(g)] remains small only in the FBC system, whereas it opens up in the SSA and OSA systems reaching the sizes larger than AC1. In the OSA system, there is a slight asymmetry between negative and positive wave vector range. In the case of AC3 [Fig.~\ref{fig:dispn}(h)], it opens up for all the considered cases. The most interesting case is present for OSA system. In the range of negative wave vectors, this AC is large, whereas in the range of positive wave vectors, it is very small, measuring only 24.67 MHz. AC4 is much larger than lower order ACs, having a size above 1 GHz, however, the size is very similar in all of the systems [Fig.~\ref{fig:dispn}(i)]. AC5 presents a similar case, with its size being even larger, measuring above 5 GHz.

\subsection{Mode profiles}

\begin{figure}[t]
    \includegraphics{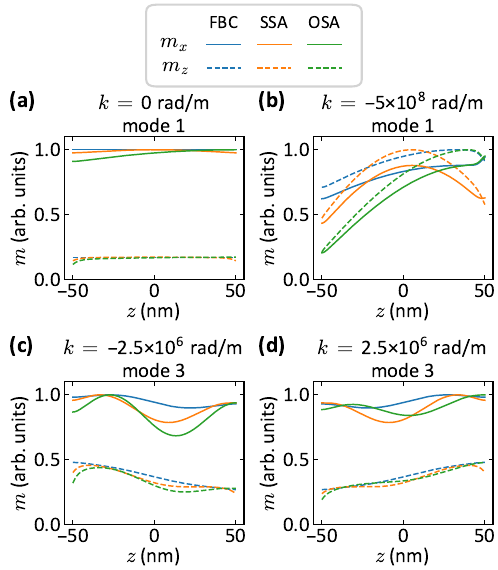}
    \caption{Distribution across the film thickness of a dynamic magnetization components $m_x$ (solid lines) and $m_z$ (dashed lines) for (a) a first mode for wave vector $k=0$, (b) a first mode for wave vector $k=\SI{-5e8}{\radian /\meter}$, (c) a third mode for wave vector $k=\SI{-2.5e6}{\radian /\meter}$, and (d) a third mode for wave vector $k=\SI{2.5e6}{\radian /\meter}$. Mode profiles are presented for system with FBC (blue lines), SSA (orange lines), and OSA (green lines). Plots present numerical simulations results only which are in agreement with analytical results.}
    \label{fig:profiles}
\end{figure}

\begin{figure*}[t]
    \includegraphics{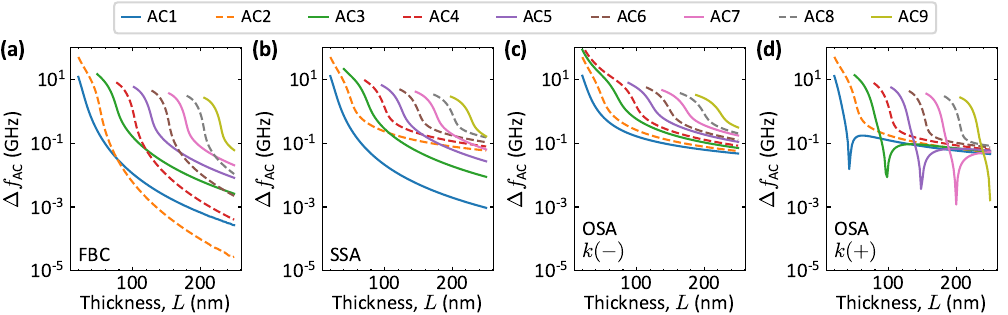}
    \caption{The AC size $\Delta f_\mathrm{AC}$ as a function of film thickness $L$ for the system with (a) FBC, (b) SSA, and OSA for (c) negative and (d) positive wave vector $k$. Odd-numbered ACs are shown with solid lines, even-numbered ACs with dashed lines. The $y$-axis is in the logarithmic scale. Plots present results of numerical simulations.}
    \label{fig:sweep-thickness}
\end{figure*}

The surface anisotropy has a significant impact on the dynamic magnetization distribution of SW modes, with mode profiles shown in Fig.~\ref{fig:profiles}. Firstly, we present the profile of the lowest frequency mode at $k=0$ in Fig.~\ref{fig:profiles}(a). Due to the low external field, the spin precession is strongly elliptical with the domination of in-plane $m_x$ component. In the case of FBC (blue lines), the mode is uniform throughout the thickness. The negative surface anisotropy leads to the reduction of the SW amplitude close to the film boundary. The mode is symmetrical for SSA, while for OSA it becomes asymmetrical. Interestingly, although the surface anisotropy affects directly only the out-of-plane $m_z$ component, the in-plane $m_x$ component is also impacted. However, in the dipole-dominated low-wave vector regime, the effect of surface anisotropy is generally small. The impact on the PSSW modes (not shown here) is even smaller. However, the anisotropy has a substantial effect on the mode profiles in the exchange-dominated large-wave vector region, as evidenced in Fig.~\ref{fig:profiles}(b) for the lowest frequency mode at $k=\SI{-5e8}{\radian /\meter}$. In both the SSA and OSA cases, the mode amplitude is significantly lower near the boundary with surface anisotropy in comparison to the FBC case. Interestingly, in this case, the $m_z$ component exceeds the $m_x$ component, and the precession is close to circular.

In Figs.~\ref{fig:profiles}(c,d), profiles of the third lowest mode are shown at $k$ between AC2 and AC3 for the negative [$k=\SI{-2.5e6}{\radian /\meter}$, Fig.~\ref{fig:profiles}(c)] and positive [$k=\SI{2.5e6}{\radian /\meter}$, Fig.~\ref{fig:profiles}(d)] wave vectors. The mode has a character of a DE mode, although the first and second term of Eq.~(\ref{eq:profile}) connected with wave vector $q_1$ also have a significant impact on the mode shape, which results in the sinusoidal character of these profiles. Their contribution is enhanced when the surface anisotropy is present. The $m_x$ component is larger than $m_z$ component, but the precession is less elliptical than when $k=0$. For both FBC and SSA cases, where the boundary conditions are identical on both surfaces, the mode profiles for opposite wave vectors are their mirror images. However, this is not true for OSA as the mode profiles differ between negative and positive wave vectors. For negative wave vectors [Fig.~\ref{fig:profiles}(c)], the contribution from first and second terms in Eq.~(\ref{eq:profile}) are significantly stronger for both $m_x$ and $m_z$ components.

\subsection{Analysis of thickness dependence}

In the next step, we present a detailed analysis of the impact of the surface anisotropy on AC formation. Firstly, we study the effect of the film thickness $L$ on the AC size $\Delta f_\mathrm{AC}$ in four cases---FBC [Fig.~\ref{fig:sweep-thickness}(a)], SSA [Fig.~\ref{fig:sweep-thickness}(b)], and OSA for both negative [Fig.~\ref{fig:sweep-thickness}(c)] and positive [Fig.~\ref{fig:sweep-thickness}(d)] wave vector $k$. In general, the increase of film thickness results in an increase in the number of ACs. This phenomenon is well-explained by the frequency limit criterion. The thickness has no impact on the maximum DE frequency $f_{k\to\infty}^\mathrm{DE}$ [Eq.~(\ref{eq:flcDE})]. In contrast, the formula for the PSSW frequency $f_{k=0}^n$ [Eq.~(\ref{eq:flcPSSW})] includes thickness in the denominator; thus, an increase of thickness results in a decrease of frequency. This allows for a higher number of PSSW modes to satisfy the frequency limit criterion, resulting in more ACs. Another relevant effect is that the AC size decreases with an increase of thickness.

Figure~\ref{fig:sweep-thickness} shows that the rate of the AC size decrease depends on the boundary conditions and the parity of the AC number. In the FBC system [Fig.~\ref{fig:sweep-thickness}(a)], the AC size decreases rapidly, but much faster for even-numbered ACs than for odd-numbered ACs. In the case of SSA [Fig.~\ref{fig:sweep-thickness}(b)], the rate of decrease for odd-numbered ACs is slightly smaller, but for even-numbered ACs the change is significant; in this case, the decrease is much smaller compared to the odd-numbered ACs. In the OSA system [Figs.~\ref{fig:sweep-thickness}(c,d)], the rate of decrease is similar across all ACs and comparable to the even-numbered ACs in the SSA system. This effect, which depends on parity, originates from the symmetry of modes and boundary conditions. Due to the dominant contribution of the $k$-dependent term in the magnetization profile shape, the DE mode has a symmetry closer to the odd-numbered PSSW modes, which are connected with the odd-numbered ACs. In the case of FBC, there is no additional source of symmetry breaking and, therefore, odd-numbered ACs are larger. On the other hand, SSA causes a symmetric disturbance of all modes, primarily affecting the dynamic magnetization amplitude in close proximity to the surface. Odd-numbered PSSW modes have opposite amplitude on the opposite boundaries, therefore, the effect of the anisotropy on the mode symmetry cancels out. On the other hand, both DE mode and even-numbered PSSW modes exhibit the same amplitude on the opposite surfaces, so the anisotropy breaks the symmetry of these modes and, as an effect, these modes induce larger ACs. In the case of OSA, the asymmetry of the anisotropy generates the asymmetry in the mode profiles, leading to large ACs in all cases. An explanation based on a simplified model of mode profiles is provided in Appendix~\ref{app2}.

The final effect is present only in the OSA system in the positive wave vector $k$ range [Fig.~\ref{fig:sweep-thickness}(d)]. It is the presence of a local minimum of AC size with a change of thickness. Interestingly, this effect only occurs for odd-numbered ACs. Upon analyzing this effect, one may question whether this local minimum reaches zero, or in other words, whether exists such a critical film thickness for which AC does not occur, i.e., the mode crossing is present. Obviously, the numerical study of the AC size can not provide a definite answer while we were not able to derive it from the analytical model. Nevertheless, the mode profiles analysis can resolve this issue.

\begin{figure}[t]
    \includegraphics{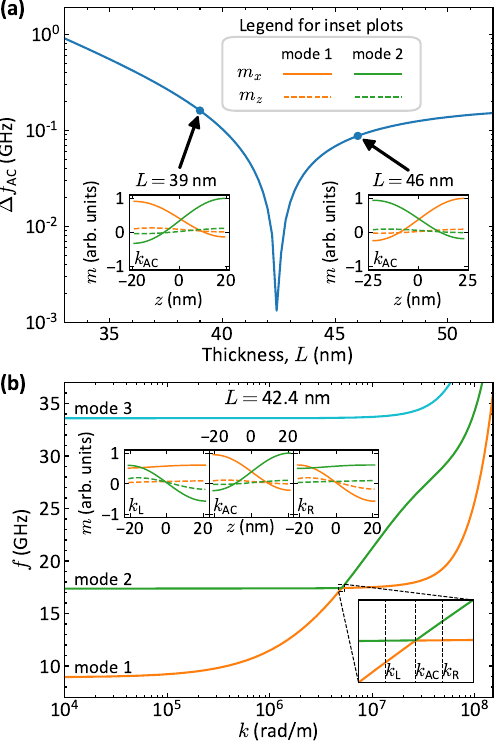}
    \caption{(a) The AC1 size $\Delta f_\mathrm{AC1}$ as a function of film thickness $L$ for the system with OSA for positive wave vector $k$---the close-up of Fig.~\ref{fig:sweep-thickness}(d) to the AC1 local minimum in small thickness range. Inset plots show the magnetization profiles of first (orange line) and second (green line) mode at $k_\mathrm{AC1}$ for the thickness of 39 nm (on the left) and 46 nm (on the right). (b) Dispersion relation of three lowest modes for the system with OSA for positive wave vector $k$ for the thickness of 42.4 nm. Inset in the bottom-right corner: the close-up to the AC1 with marking of three wave vectors -- $k_\mathrm{L}=\SI{4.853e6}{\radian /\meter}$, $k_\mathrm{AC}=\SI{5.033e6}{\radian /\meter}$, and $k_\mathrm{R}=\SI{5.2e6}{\radian /\meter}$. Inset at top: magnetization profiles of first (orange line) and second (green line) mode at $k_\mathrm{L}$ (left), $k_\mathrm{AC}$ (center), and $k_\mathrm{R}$ (right). Plots present results of numerical simulations.}
    \label{fig:profac}
\end{figure}

The close-up to the local minimum of the AC1 [Fig.~\ref{fig:sweep-thickness}(d), solid blue line] is shown in Fig.~\ref{fig:profac}(a). In this case, the step in simulation was 0.2 nm. The minimum value of $\Delta f_\mathrm{AC1} = 1.33$ MHz was obtained for a thickness of 42.4 nm. We can take a look on the magnetization profiles for the first and second modes at wave vector $k_\mathrm{AC}$ [the inset plots in Fig.~\ref{fig:profac}(a)] for the thickness smaller (39 nm, left plot) and larger (46 nm, right plot) than the thickness of the AC1 size minimum. In both cases, the mode profiles are very similar, indicating the superposition of the DE and first PSSW mode. However, the most important fact is to notice that the modes are interchanged. For $L = \SI{39}{\nano\meter}$, the lower frequency mode (orange line) has higher amplitude at the bottom of the film, while for $L = \SI{46}{\nano\meter}$, higher amplitude is at the top of the film. The higher frequency mode (green line) demonstrates the opposite trend. The detailed analysis of the profiles indicates that this behavior is connected with each local minimum of $\Delta f_\mathrm{AC}$, i.e. the mode profiles at $k_\mathrm{AC}$ interchange. According to our analysis, it indicates the existence of a critical thickness value $L_\mathrm{C}$ where a crossing occurs instead of AC, indicating the absence of a gap between the first and second mode. This observation suggests the possible occurrence of an accidental degeneracy in the system \cite{Her37,Hua11}, meaning that there are two solutions with the same values of wave vector and frequency.

\begin{figure*}[t]
    \includegraphics{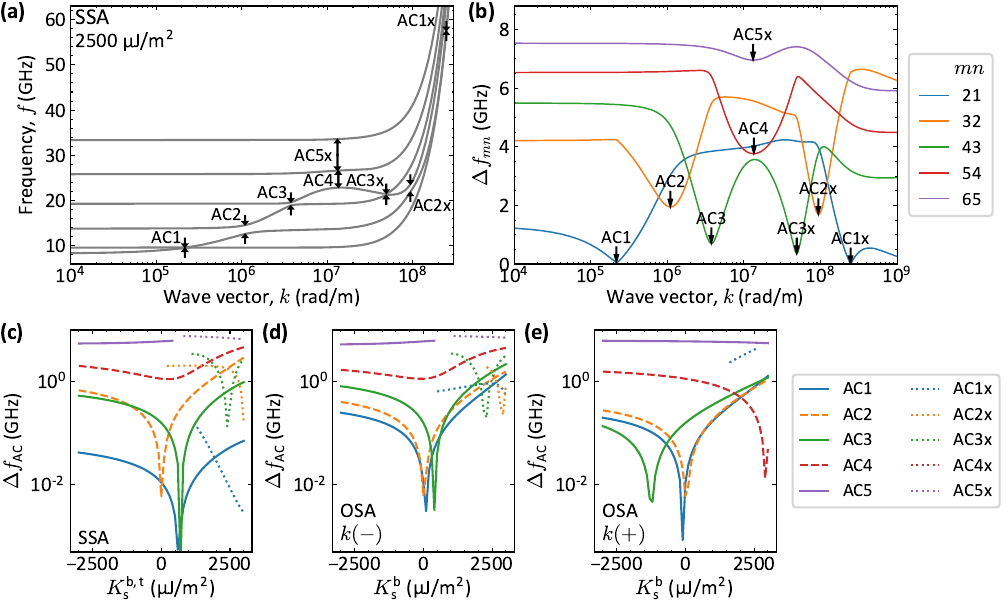}
    \caption{(a) Dispersion relation of the lowest six modes of the system with SSA for $K_\mathrm{s}=\SI{2500}{\micro \joule /\meter^2}$. (b) Frequency difference between the neighboring modes of the system with SSA presented in (a). The $x$-axis is in the logarithmic scale. (c-e) The AC size $\Delta f_\mathrm{AC}$ as a function of the surface anisotropy $K_\mathrm{s}$ for the system with (c) SSA and OSA for (d) negative and (e) positive wave vector $k$. Odd-numbered ACs are shown with solid lines, even-numbered ACs with dashed lines, and second-order ACs with dotted lines. The $y$-axis is in the logarithmic scale. Plots present results of numerical simulations.}
    \label{fig:sweep-anis}
\end{figure*}

Another observation concerns the system with the lowest value of $\Delta f_\mathrm{AC1}$ found for the thickness of 42.4 nm. Its dispersion relation is shown in Fig.~\ref{fig:profac}(b). The AC1 is not visible in the full dispersion. Interestingly, the AC1 is still too small to be visible even after a close-up of the AC1 vicinity (inset plot in the lower right corner). To study the mode profiles in the vicinity of AC1, we chose three wave vectors: $k_\mathrm{AC}=\SI{5.033e6}{\radian /\meter}$, $k_\mathrm{L}=\SI{4.853e6}{\radian /\meter}$, and $k_\mathrm{R}=\SI{5.2e6}{\radian /\meter}$. The mode profiles are shown in the inset plot at the top part of Fig.~\ref{fig:profac}(b). For $k_\mathrm{AC}$ (middle plot), the profiles are similar to the case of $L = \SI{39}{\nano\meter}$. It suggests, that the critical value of the thickness $L_\mathrm{C} > \SI{42.4}{\nano\meter}$. In the case of $k_\mathrm{L}$ (left plot), the profile of the first mode (orange line) has a character of the DE mode with a small amplitude reduction at the bottom due to the surface anisotropy. The second mode (green line) has the character of the first PSSW mode. This mode has a slightly larger amplitude at the bottom than at the top. The modes at $k_\mathrm{R}$ (right plot) have the same character as the modes at $k_\mathrm{L}$, but their order is reversed. It clearly shows that far from the AC (where $f_2-f_1 \gg \Delta f_\mathrm{AC1}$), the modes have the same character on both sides of the AC, as if the interaction between them is negligible. It is worth noting that this interchange is not so clear in the case where $\Delta f_\mathrm{AC}$ is relatively large. In this case, the intermixing of the effects of the wave vector dependence and the short distance between the ACs relative to their size leads to a significant change in the mode profiles.

\subsection{Analysis of surface anisotropy constant dependence}

The analysis presented above was done for the case where the surface anisotropy constant $K_\mathrm{s}$ has a negative value, resulting in the partial pinning condition for the out-of-plane dynamic component of the magnetization. Now we can look at the case where $K_\mathrm{s}$ is positive, so the magnetization amplitude close to the surface is enhanced. The dispersion relation for the system with SSA for $K_\mathrm{s}^\mathrm{t} = K_\mathrm{s}^\mathrm{b} = \SI{2500}{\micro \joule /\meter^2}$ is shown in Fig.~\ref{fig:sweep-anis}(a). The small $k$ range is comparable to the case of negative $K_\mathrm{s}$. However, for about $k = 10^7$ rad/m, the DE mode reaches the local maximum at about 23 GHz and obtains negative group velocity. This effect is analogous to the effect of the volume perpendicular magnetic anisotropy \cite{Ban17}. On its way, the DE mode produces additional ACs, which did not occur in the case of FBC and negative $K_\mathrm{s}$. The frequency difference between the adjacent modes [Fig.~\ref{fig:sweep-anis}(b)] shows that additional ACs are present for the first, second, and third PSSW modes. These ACs are marked with the letter 'x'. Also, an AC5 is present. However, it is not related to the AC5 occurring for negative anisotropy, therefore, it is also marked with 'x'.

\begin{figure*}[t]
    \includegraphics{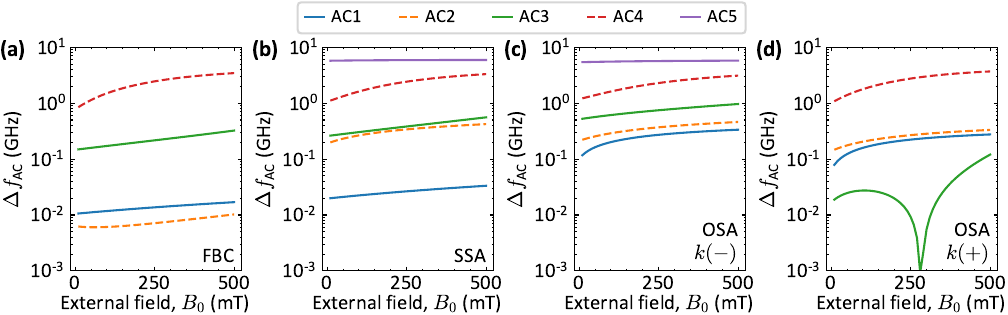}
    \caption{The AC size $\Delta f_\mathrm{AC}$ as a function of the external magnetic field $B_0$ for the system with (a) FBC, (b) SSA, and OSA for (c) negative and (d) positive wave vector $k$. Odd-numbered ACs are shown with solid lines, even-numbered ACs with dashed lines. The $y$-axis is in the logarithmic scale. Plots present results of numerical simulations.}
    \label{fig:sweep-field}
\end{figure*}

Next, we study the AC size as a function of the surface anisotropy constant $K_\mathrm{s}$ for the case of SSA [Fig.~\ref{fig:sweep-anis}(c)] and OSA for negative [Fig.~\ref{fig:sweep-anis}(d)] and positive [Fig.~\ref{fig:sweep-anis}(e)] wave vector $k$. We calculated it numerically in the range from $\SI{-3000}{\micro \joule /\meter^2}$ to $\SI{3000}{\micro \joule /\meter^2}$ with a step of $\SI{100}{\micro \joule /\meter^2}$. Almost all curves have a minimum similar to the one present in Fig.~\ref{fig:sweep-thickness}(d). A detailed analysis of the mode profiles agrees with the previous observation---in each case, the mode profiles at $k_\mathrm{AC}$ interchange, so we expect that for a critical value of $K_\mathrm{s}$ a crossing between modes should occur. The position of the minimum depends on the AC parity. AC2 has the smallest size at $K_\mathrm{s} = 0$ (however, we expect the critical value to be very low, i.e., $|K_\mathrm{s}^\mathrm{critical}| < \SI{50}{\micro \joule /\meter^2}$). The odd-numbered ACs (AC1 and AC3) have the smallest size for positive $K_\mathrm{s}$ in the system with SSA and OSA at negative $k$, while for the system with OSA at positive $k$, the smallest value occurs for negative $K_\mathrm{s}$. Interestingly, for the system with OSA, $K_\mathrm{s}^\mathrm{critical}$ of the same AC is different for positive and negative wave vector range, which means that we can get a situation where the AC is present only on one side of the dispersion relation, while on the opposite side a crossing will be present. In general, the AC tends to have larger size for positive surface anisotropy than for negative surface anisotropy of the same value. In addition, we can see that for a wide range of positive surface anisotropy, additional ACs (marked with the letter 'x') occur in all systems. Their source lies in the negative slope range of the dispersion relation, as discussed above. Their minima also follow the rule of the mode profile interchange, so we should expect these minima to go to zero as well.

\subsection{Analysis of external magnetic field dependence}

Finally, the effect of the external magnetic field $B_0$ on the AC size is shown in Fig.~\ref{fig:sweep-field}. The ACs have been calculated in the field range between 10 and 500 mT with a step of 10 mT. Almost all ACs are increasing with the increase of the external field. This observation correlates with the fact that $k_\mathrm{AC}$ also increases with the increase of external field. Then, the $k$-dependent terms in the DE mode profile [Eq.~(\ref{eq:profile})] give a stronger contribution, and the profile asymmetry is increased, resulting in a stronger interaction with PSSW modes and a larger AC size. The most remarkable example is AC4 in the FBC system, which increases by a factor of 4.06 in the investigated field range. On the other hand, AC5 in the SSA system increases by only 3\% in the same range. Interestingly, in the OSA system, the local minimum for the AC3 occurs in the positive wave vector range. The lowest detected value is 0.94 MHz at 280 mT. This minimum also has the source in the mode profile interchange at $k_\mathrm{AC}$, indicating the closing of the AC gap. In the direction of lower fields, the local maximum is present for 100 mT with the AC size of 27.4 MHz, while for higher fields it increases up to 120 MHz in the upper limit of the study of 500 mT. The results show that the external field provides a simple way to control the AC size, which is the easiest source of control from the experimental point of view. 


\section{Conclusions}

In this article, we provide a comprehensive investigation of the SW dynamics in the ferromagnetic film in the DE geometry in the presence of surface anisotropy with the use of analytical model and numerical simulations. We compare three different cases: free boundary conditions, symmetrical surface anisotropy, and one-sided surface anisotropy. We show that the surface anisotropy significantly increases the size of the AC between DE and PSSW modes. In the case of OSA, the mirror symmetry breaking leads to the asymmetrical dispersion relation with respect to the wave vector $k$, which particularly affects the AC size. The surface anisotropy also has a strong influence on the shape of the mode profiles. 

We have studied in detail the impact of various parameters (i.e., film thickness, surface anisotropy constant, and external magnetic field) on the AC size. In general, the ACs shrink with the increase of film thickness or the decrease of the external magnetic field. Also, the parity of the AC has a strong influence on the AC size. For large positive surface anisotropy constant, the mode of DE character has a non-monotonic dispersion relation, which leads to the appearance of additional ACs for large wave vectors. In most cases, the increase in anisotropy leads to the increase in AC size. Interestingly, we found that under certain conditions, the AC can close and turn into a crossing. This phenomenon, known as accidental degeneracy, occurs for some particular ACs when the value of the surface anisotropy constant, the layer thickness, or the external magnetic field is changed. In the system with SSA, it occurs for both negative and positive wave vector, while in the system with OSA only on one side of the dispersion relation for a given set of parameters. The transition through the accidental degeneracy point in any parameter space is always associated with the exchange of the order of the mode profiles in the AC region. It is worth to note that the results shown in the paper are calculated for typical material parameters of CoFeB but the presented effects are universal and should also occur for different materials.

The presence of surface anisotropy in magnetic thin films is ubiquitous. It is often considered a detrimental feature, but it can also be an essential property. The ability to control the anisotropy by voltage as well as to control its effects by an external magnetic field gives an additional advantage. Moreover, surface anisotropy of different strength on opposite surfaces provides a simple way to induce the nonreciprocity in the structure. We believe that surface anisotropy can be exploited in magnonic devices where asymmetrical transmission or the possibility to control the propagation of the SW is a fundamental property.


\begin{acknowledgments}
K.S. and M.K. acknowledge the financial support from National Science Centre, Poland, grants no. UMO-2020/39/I/ST3/02413 and UMO-2021/41/N/ST3/04478. The research leading to these results has received funding from the Norwegian Financial Mechanism 2014-2021 project no. UMO-2020/37/K/ST3/02450. K.S. acknowledges the financial support from the Foundation for Polish Science (FNP).

The dataset for this manuscript is available on https://doi.org/10.5281/zenodo.8382924.
\end{acknowledgments}


\appendix
\section{Relation between wave vectors $k$ and $q_1$}\label{app1}
\begin{figure}[t]
    \includegraphics{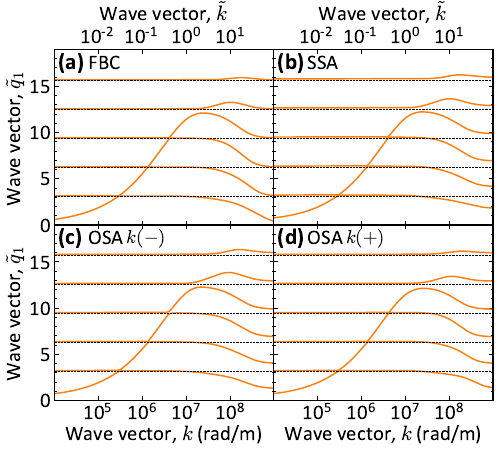}
    \caption{Wave vector $\Tilde{q}_1=q_1 L$ as a function of wave vector $k$ of six lowest modes of a CoFeB film of thickness $L=\SI{100}{\nano \meter}$ with (a) FBC, (b) SSA with $K_\mathrm{s}^\mathrm{t}=K_\mathrm{s}^\mathrm{b}=\SI{-700}{\micro \joule /\meter^2}$, and (c,d) OSA with $K_\mathrm{s}^\mathrm{t}=0$ and $K_\mathrm{s}^\mathrm{b}=\SI{-1500}{\micro \joule /\meter^2}$ for (c) negative and (d) positive wave vector $k$ in the external magnetic field $\mu_0 H_0 = \SI{50}{\milli \tesla}$. Horizontal dashed black lines represent the values $q_1=n\pi/L$ for $n=1,2,3...$ Plots present analytical results.}
    \label{fig:q1k}
\end{figure}

Figure~\ref{fig:q1k} shows the wave vector $\Tilde{q}_1 = q_1 L$ as a function of wave vector $k$ for three cases studied in the manuscript---FBC [Fig.~\ref{fig:q1k}(a)], SSA with $K_\mathrm{s}^\mathrm{t}=K_\mathrm{s}^\mathrm{b}=\SI{-700}{\micro \joule /\meter^2}$ [Fig.~\ref{fig:q1k}(b)], and OSA with $K_\mathrm{s}^\mathrm{t}=0$ and $K_\mathrm{s}^\mathrm{b}=\SI{-1500}{\micro \joule /\meter^2}$ for negative [Fig.~\ref{fig:q1k}(c)] and positive [Fig.~\ref{fig:q1k}(d)] wave vector $k$. In the low wave vector range (up to about $10^7$ rad/m), the plots are very similar to the dispersion relations shown in Figs.~\ref{fig:dispn}(a-d), including the presence of the gaps between the modes. For the case of FBC, the PSSW modes are placed exactly at $q_1=n\pi /L$, while for DE mode the value of $q_1$ is increasing and produces ACs with PSSW modes exactly as in the dispersion relation. Nevertheless, the large value of $q_1$ for DE mode is not decisive for the shape of the mode profile since the coefficients in Eq.~(\ref{eq:profile}) associated with $q_1$ give smaller contribution that those associated with $k$ (however, this contribution is not negligible). For the case of SSA and OSA, values of $q_1$ of PSSW modes are larger than $n\pi /L$. It is clear that a larger value of $q_1$ results in larger frequency of PSSW modes according to Eq.~(\ref{eq:omega}).

In the large $k$ range for the case of FBC, the values of $q_1$ go back to $n\pi /L$, but this time for $n$ starting from 0. To achieve this feat, all modes in the range of the DE mode are decreasing in the value of $q_1$ in the dipole-exchange regime of the wave vector $k$. In the case of SSA and OSA, the values of $q_1$ are also larger than $n\pi /L$ but the difference is much larger than in the small $k$ range.

\section{Toy model of interaction between modes}\label{app2}
\begin{figure*}[t]
    \includegraphics{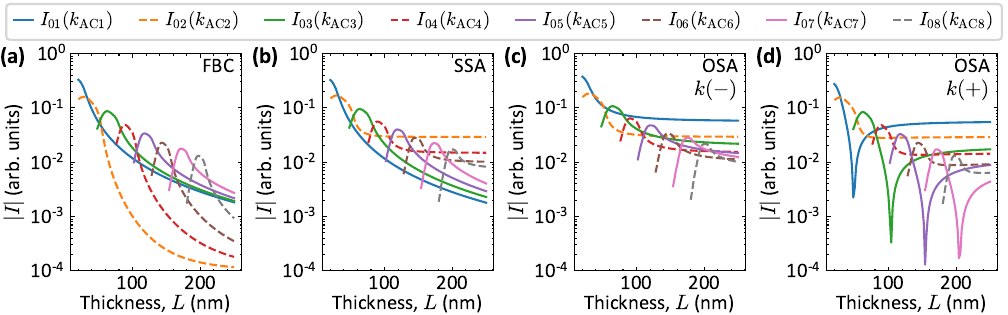}
    \caption{The absolute value of overlapping integral $I$ as a function of film thickness $L$ for the system with (a) FBC, (b) SSA, and OSA for (c) negative and (d) positive wave vector $k$. Odd-numbered ACs are shown with solid lines, even-numbered ACs with dashed lines. The $y$-axis is in the logarithmic scale.}
    \label{fig:oi}
\end{figure*}

The interaction between the modes can be explained using of a simplified model of mode profiles. In the case of FBC at $k=0$, the modes form a basis of cosine functions
\begin{equation}\label{eq:mn}
    m_n = A_n \cos{\left(n\pi \left(z-\frac{L}{2}\right)\right)},
\end{equation}
where $m_0$ represents the DE mode and $m_{n>0}$ represents $n$-th order PSSW modes. $A_n$ is the normalization constant which assure that $\int_{-L/2}^{L/2} m_n^2 \mathrm{d}z = 1$.

Assume that in the regime of small $k$, the PSSW modes remain unchanged with the change of the wave vector $k$, so their profiles are represented by Eq.~(\ref{eq:mn}). On the other hand, the DE mode is described by the function
\begin{equation}
    m_0(k) = A_0 e^{kz}.
\end{equation}

In the presence of negative surface anisotropy, the PSSW modes are “squeezed” to satisfy the boundary conditions. Due to this effect, their profiles are modified such that $q_1 = (n+p_n)\pi /L$, where $p_n$ is the relative shift of wave vector. In the case of SSA, the mode profile is modified in the following way:
\begin{equation}
    m_n = A_n \cos{\left((n+p_n)\pi \left(z-\frac{n}{n+p_n}\frac{L}{2}\right)\right)}.
\end{equation}
In the case of OSA, the mode profile of PSSW modes is represented by the function:
\begin{equation}
    m_n = A_n \cos{\left((n+p_n)\pi \left(z-\frac{L}{2}\right)\right)}.
\end{equation}
We assume that the change in DE mode due to surface anisotropy is negligible. Basing on the results in Fig.~\ref{fig:q1k}, we assume that $p_n$ has a constant value of 0.03 for all PSSW modes and all thicknesses.

Strength of the interaction between the modes is described by the overlapping integral
\begin{equation}
    I_{ij} = \int_{-L/2}^{L/2} m_i m_j \mathrm{d}z.
\end{equation}

Figure~\ref{fig:oi} shows the overlapping integral between the DE mode and the $n$-th order PSSW mode at $k_{\mathrm{AC}_n}$ as a function of layer thickness $L$. The model qualitatively reproduces the behavior shown in Fig.~\ref{fig:sweep-thickness} showing that the overlapping integral is connected with the AC size. Firstly, there is an identical dependence on the PSSW mode parity. In the case of FBC [Fig.~\ref{fig:oi}(a)], the overlapping integral has a larger value for the function representing odd-numbered PSSW modes than for even-numbered PSSW modes. In the case of SSA [Fig.~\ref{fig:oi}(b)], the overlapping integral for even-numbered PSSW modes grows over the integral for odd-numbered PSSW modes which only increases slightly compared to the FBC case. In the case of OSA for negative wave vectors $k$ [Fig.~\ref{fig:oi}(c)], the value of the overlapping integral is similar for all modes. In the case of positive wave vectors $k$ [Fig.~\ref{fig:oi}(d)], we have successfully reproduced the presence of the minima for odd-numbered PSSW modes shown in Fig.~\ref{fig:sweep-thickness}(d). The positions of the minima---at 48, 104, 154, and 204 nm for the first, third, fifth, and seventh PSSW modes, respectively, is in good agreement with the positions of the minima in Fig.~\ref{fig:sweep-thickness}(d) (42, 98, 148, and 200 nm, respectively).

\end{document}